\documentclass[aps,prb,showpacs,twocolumn]{revtex4}%
\usepackage{amssymb}
\usepackage{amsmath}
\usepackage{graphicx}
\usepackage{dcolumn}
\usepackage{bm}
\usepackage{units}
\usepackage{color}
\usepackage{amsfonts}%
\providecommand{\U}[1]{\protect\rule{.1in}{.1in}}
\begin{document}
\title{Angular momentum transfer torques in spin valves with perpendicular magnetization}
\author{Xingtao Jia}
\affiliation{Department of Physics, Beijing Normal University, Beijing 100875, China}
\author{Ying Li}
\affiliation{Department of Physics, Beijing Normal University, Beijing 100875, China}
\author{Ke Xia}
\affiliation{Department of Physics, Beijing Normal University, Beijing 100875, China}
\author{Gerrit E. W. Bauer}
\affiliation{Delft University of Technology, Kavli Institute of NanoScience, 2628 CJ Delft,
The Netherlands}
\affiliation{Institute for Materials Research, Tohoku University, Sendai 980-8577, Japan}
\date{\today }

\begin{abstract}
Spin valves incorporating perpendicularly magnetized materials are promising
structures for memory elements and high-frequency generators. We report the
angular dependence of the spin-transfer torque in spin valves with
perpendicular equilibrium magnetization computed by first-principles circuit
theory and compare results with experiments by W.H. Rippard \textit{c.s}.
[Phys. Rev. B\textbf{\ 81}, 014426 (2010)] on the CoFe$|$Cu$|$CoNi system.
Furthermore, we predict a non-monotonous ("wavy") spin-transfer torque when
the Cu spacer is replaced by a Ru layer.

\end{abstract}

\maketitle


\section{Introduction}

A current can be used to read out the information in magnetic memory devices
by the giant magnetoresistance. Magnetic random access memory (MRAM)
technology has become scalable by writing information using the
current-induced spin-transfer torques
(STT).\cite{Slonc96,Berger96,Myers99,dwr} The critical electric current
density $j_{c}$ necessary to switch a magnetic layer in a spin-valve structure
is an all-important figure of merit in this case. The introduction of
materials with perpendicular magnetocrystalline anisotropies that forces the
equilibrium magnetization out of the plane,\cite{kent04} has helped to reduce
$j_{c}$.\cite{Mangin06,Mangin09,Yoda}

Co$|$Ni multilayers are an interesting system with perpendicular
anisotropy,\cite{Daal92,Mangin06,sun08} with a higher polarization and less
spin-flip scattering than, for example, CoPt alloy.\cite{Inomata98} Rippard
\textit{et al}.\cite{Ripp2010} studied current-induced high-frequency
generation in structures with a perpendicularly polarized (Co$|$Ni)$_{n}$
multilayer serving as the switchable magnet and an in-plane magnetized Co
layer as polarizer. The output power of such a device depends sensitively on
the asymmetry of the angular dependent STT when the magnetization of the free
layer is reversed.\cite{Slonc96,dwr,slon2002} By generating an rf output by a
dc current in a spin valve in which the free layer is magnetized normal to the
polarizing layer, Rippard\textit{\ c.s}. parameterized the skewness of the
torque as a function of magnetization angle.\cite{Ripp2010} Koyama \textit{et
al.} \cite{Tomohiro2008} measured high speed current-induced domain wall
velocities (40 $%
\operatorname{m}%
/%
\operatorname{s}%
$) in magnetic perpendicular Co$|$Ni multilayers with current-in-plane
configuration. Another interesting materials system is Co$|$%
Ru,\cite{Parkin90,song2006} which also displays perpendicular magnetic
anisotropy.\cite{Miyawaki2009}

Semiclassical theories\cite{ab00,zang202} that combine a quantum treatment of
the interface scattering and diffusion treatment of bulk scattering in general
explain experiments on magnetic metallic multilayers well.\cite{dwr} Here we
report calculations of the STT of spin valves containing perpendicularly
oriented ferromagnetic materials based on magnetoelectronics circuit theory
using interface transport parameters computed by first principles. The
spin-orbit coupling is the origin of the magnetic crystalline anisotropy and
perpendicular magnetization. However, the experimental spin-dependent
interface resistances for not too heavy elements can be reproduced by
parameter-free calculations without taking into account the spin orbit
interaction,\cite{Bass10} which will therefore be disregarded in the following.

Here, we study the angular dependent STT in Co$_{1}$Ni$_{x}$ (the subscripts
refer to the number of atomic layers) based spin valves by circuit theory in
combination with first-principles calculations. Firstly, we present results
for Co$|$Cu$|$(Co$_{1}$Ni$_{x}$)$_{y}$Co$_{1}|$Cu(111) stuctures, where the
subscripts $1$ and $x$ indicate again the number of atomic layers, while $y$
is the number of stacks and compare them with experiments.\cite{Ripp2010}
Next, we report large and `wavy' angular-dependent STT for Co$|$Ru$|$(Co$_{1}%
$Ni$_{2}$)$_{x}$Co$_{1}|$Ru(111) spin valves which might therefore be very
efficient high-frequency generators.

In Sec.~\ref{sec:Method}, we introduce our method to calculate the STT in spin
valves in terms of the spin mixing conductances of the interfaces computed
from first principle, including corrections for the magnetically active bulk
material and the diffusive environment. In Sec.~\ref{sec:discussion} we
present results for the spin mixing conductances for two the two types of spin
vales with perpendicular magnetic anisotropy and compute the angular
dependence of STT by magnetoelectronic circuit theory. We summarize our
results in Sec.~\ref{sec:sum}.

\section{Spin mixing conductance in a diffusive environment}

\label{sec:Method} 
\begin{figure}[ptb]
\includegraphics[width=8.6cm]{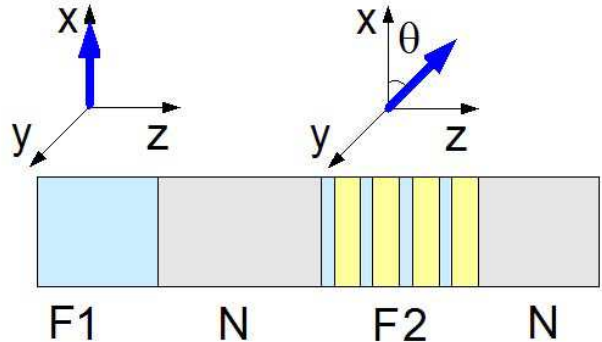}\ \caption{Scheme of asymmetric
F1$|$N$|$F2$|$N spin valves with perpendicular magnetization F2 used in the
calculations }%
\label{scheme}%
\end{figure}

The STT due to a current bias $I$ in ferromagnet$|$%
normal-metal$|$ferromagnet (F$|$N$|$F) spin valves in which the magnetizations
are at an angle $\theta$ can be computed analytically by circuit
theory\cite{Alex2002,dwr} and, assuming structural symmetry, be parameterized
as\cite{slon2002}%
\begin{equation}
\tau(\theta)=\frac{\hbar I\tilde{P}}{4e}\frac{\Lambda\sin\theta}{\Lambda
\cos^{2}(\theta/2)+\Lambda^{-1}\sin^{2}(\theta/2)},\label{TORQUE}%
\end{equation}
where the asymmetry parameter can be expressed in terms of the the parameters
of the N$|$F interface as $\Lambda=|\tilde{\eta}|/\sqrt{\left(  1-\tilde
{P}^{2}\right)
\operatorname{Re}%
\tilde{\eta}}$, where $\tilde{\eta}=2\tilde{G}_{\uparrow\downarrow}/\left(
\tilde{G}_{\uparrow}+\tilde{G}_{\downarrow}\right)  $ is the normalized
effective spin-mixing conductance and $\tilde{P}=\left(  \tilde{G}_{\uparrow
}-\tilde{G}_{\downarrow}\right)  /\left(  \tilde{G}_{\uparrow}+\tilde
{G}_{\downarrow}\right)  $ is the conductance polarization. Here, $\tilde
{G}_{\uparrow},\tilde{G}_{\downarrow}$ and $\tilde{G}_{\uparrow\downarrow}$
are the spin-dependent and spin-mixing conductances, respectively, where the
tilde indicates that they have been \textquotedblleft Schep
corrected\textquotedblright\ for a diffusive environment and include the
effects of the magnetically active contact regions close to the interface. In
deriving Eq. (\ref{TORQUE}) spin flip in the normal layer has been
disregarded. When the spin-flip diffusion length in the magnetic layers is
much longer than the bulk layer thickness:\cite{dwr}
\begin{equation}
\frac{1}{\tilde{G}_{\sigma}}=\frac{1}{G_{\sigma}}+\frac{1}{2}\frac{e^{2}}%
{h}\left(  \frac{\rho_{F,\sigma}d_{F}}{A_{F}}\right)  -\frac{1}{2}\left(
\frac{1}{G_{N}^{sh}}+\frac{1}{G_{F,\sigma}^{sh}}\right)  \label{Gtilde}%
\end{equation}
and%
\begin{equation}
\frac{1}{\tilde{G}_{\uparrow\downarrow}}=\frac{1}{G_{\uparrow\downarrow}%
}-\frac{1}{2G_{N}^{sh}}.
\end{equation}
where $\sigma$ is the spin index, $d_{F(N)}$ the thickness of ferromagnet $F$
or normal metal $N$ layer, $\rho$ the bulk resistivity (for a single spin),
and $A_{F}$ the pillar cross section. The $G^{sh}$'s are Sharvin
conductances,\ $G_{\uparrow}=\left(  e^{2}/h\right)  \mathrm{\,tr\,}%
\mathbf{t}_{\uparrow}^{\dagger}\mathbf{t}_{\uparrow}$, $G_{\downarrow}=\left(
e^{2}/h\right)  \mathrm{\,tr\,}\mathbf{t}_{\downarrow}^{\dagger}%
\mathbf{t}_{\downarrow}$ and $G_{\uparrow\downarrow}=\left(  e^{2}/h\right)
\mathrm{\,tr\,}\left(  \mathbf{I}-\mathbf{r}_{\uparrow}^{\dagger}%
\mathbf{r}_{\downarrow}\right)  $, where $\mathbf{t}_{\downarrow(\downarrow
)}\ \left(  \mathbf{r}_{\uparrow(\downarrow)}\right)  $ are the matrices of
the transmission (reflection) coefficients of the phase coherent region of the
N$|$F contact as seen from the normal metal and at the Fermi energy.
$\mathbf{I}$ is an $M\times M$ unit matrix, where $M$ is the number of
conducting channels in $N$. The third term on the right-hand side of the last
two equations are the Schep correction, while the second terms correct for the
magnetically active bulk regions. When the ferromagnetic layer is much thicker
than the spin-flip diffusion length $l_{sd}^{F}$, the latter should replace
$d_{F}$ in Eq. (\ref{Gtilde}). With the spin-orbit interaction we also ignore
intrinsic spin-flip scattering at the interfaces. The ferromagnetic layers are
assumed sufficiently thick such that mixing transmission contribution may be
disregarded.\cite{Alex2006} Note that Eq. (\ref{TORQUE}) only holds for
structurally symmetric spin valves. In the following we use the general
expression in which the left and right interface parameters differ, as shown
in Fig.\ref{scheme}, but do not list the expressions explicitly here (see
Refs. \onlinecite{Alex2002,Xiao2004,Barnas2005,Rychkov2009}).

In our calculations the atomic potentials were determined in the framework of
the tight-binding (TB) linear muffin-tin- orbital (MTO) method\cite{Turekbook}
based on density functional theory in the local density approximation and an
exchange-correlation potential parameterized by von Barth and Hedin.\cite{BH}
The self-consistent crystal potentials were used as input to a TB-MTO
wave-function-matching calculation, from which we obtained the transmission
and reflection at the interfaces. The calculations are carried out with a
$\mathbf{k_{||}}$ mesh density equivalent to more than 3600 $\mathbf{k_{||}}$
mesh points in the two dimensional Brillouin zone (BZ) corresponding to the
interface unit cell. The technical detail can be found in Ref.
\onlinecite{xia06}. Table \ref{tab_parametrs} compiles our results for various
interface conductances including the bulk corrections due to magnetically
active regions.

\section{Co$_{1}$Ni$_{2}|$Cu and Co$_{1}$Ni$_{2}|$Ru multilayers}

\label{sec:discussion}

We first focus on the Co$\mathrm{|}$Ni multilayers, which we treat as phase
coherent regions, \textit{i.e}. we compute the scattering matrix of the entire
multilayers, which is then treated in the circuit theory of conventional spin
valves just like a single interface. We present the spin-dependent and mixing
conductances of $\mathrm{Cu|X}_{n}\mathrm{|Cu}$ with [$\mathrm{X}%
_{n}\mathrm{=(Co}_{1}\mathrm{Ni}_{2}\mathrm{)}_{n}\mathrm{Co}_{1}$]. Here the
Cu leads on both sides are semi-infinite. $\mathrm{X}_{n}$ denotes $n$
repetitions of the Co$_{1}$Ni$_{2}$ multilayer unit. As in the
experiments,\cite{Ripp2010} a Co atomic layer is added for better contact with
the Cu reservoirs. Since samples have been grown by sputtering, we take
interface disorder into account, which is in general well modeled by a two
monolayer 50\%-50\% interfacial alloy (Co$_{1}$Ni$_{2}$)$_{n}\rightarrow
(\left[  \mathrm{Co}_{0.5}\mathrm{Ni}_{0.5}\right]  $Ni$\left[  \mathrm{Co}%
_{0.5}\mathrm{Ni}_{0.5}\right]  $)$_{n}$.\cite{dwr} Spin-flip scattering at
the Co$|$Ni interface will suppress any benefits of an even larger number of
Co$|$Ni interfaces.\cite{Bass10} We therefore present here only calculations
with $n\leq5.$ The computed dimensionless mixing conductance $\tilde{\eta}$ is
also listed in the table.

\begin{figure}[ptb]
\caption{Comparison of computed and experimental (Ref. \onlinecite{Ripp2010})
angular dependent STT in F1$|$Cu$|$(Co$_{1}$Ni$_{2}$)$_{5}$Co$_{1}|$Cu (F1=Co,
Co$_{90}$Fe$_{10}$) spin valves with two monolayer $50\%$-$50\%$
\ \ intermixed interfaces. For Co as fixed lead, we vary the thickness
$d_{\mathrm{Co}}$ from $5\operatorname{nm}$ to $20\operatorname{nm}$. When
using Co$_{90}$Fe$_{10}$ as fixed layer, we use $d_{\mathrm{Co}_{90}%
\mathrm{Fe}_{10}}=2.5\operatorname{nm}$, and resistivity $\rho_{\mathrm{Co}%
_{90}\mathrm{Fe}_{10}}=154\operatorname{\Omega }\operatorname{nm}$ (Ref.
\onlinecite{Jiang2003}) and Co$|$Cu interface parameters. The dark area
indicated the experiment results parameterized by Slonczewski's formula with
$\Lambda=1.3 $ and $\Lambda=1.7$. Calculations are carried out by circuit
theory for an asymmetric spin valve with first-principles interface parameters
using the Schep correction including the contribution form the magnetically
active region of the bulk ferromagnet as described in the text.}%
\label{tor_comp}%
\includegraphics[width=8.6cm]{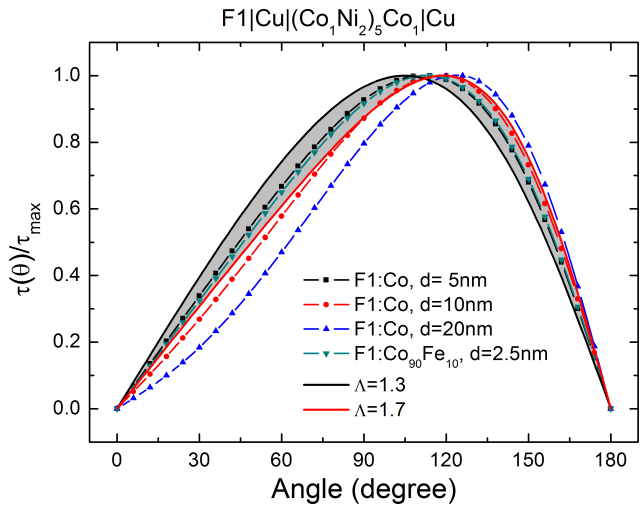}\ \end{figure}

In the fcc crystal structure Co and Ni have nearly identical band structures
for the majority spin, which results in very transparent Co$|$Ni interfaces.
The majority spin conductance therefore stays nearly constant with increasing
$n$. For minority spin electrons the scattering at the Co$|$Ni interface is
much stronger. Consequently the minority spin conductance decreases rapidly
with increasing number of Co$|$Ni interfaces.

\begin{figure}[ptb]
\includegraphics[width=8.6cm]{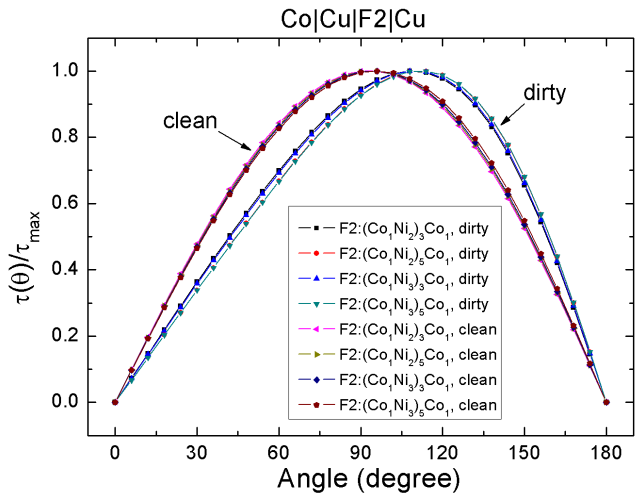}\ \caption{Angular dependent STT in
Co$|$Cu$|$(Co$_{1}$Ni$_{2}$)$_{x}$Co$_{1}|$Cu and Co$|$Cu$|$(Co$_{1}$Ni$_{3}%
$)$_{x}$Co$_{1}|$Cu spin valves with $d_{\mathrm{Co}}=5\,\operatorname{nm}$. }%
\label{coni}%
\end{figure}

Fig. \ref{tor_comp} shows the angular dependent STT exerted on the right-hand
side of F1$|$Cu$|$(Co$_{1}$Ni$_{2}$)$_{5}$Co$_{1}|$Cu ($\mathrm{F1=Co}$,
Co$_{90}$Fe$_{10}$) spin valves with intermixed interfaces calculated by our
first-principles circuit theory and compared with the experimental
result.\cite{Ripp2010} For pure Co as fixed lead, we vary $d_{\mathrm{Co}}$
from $5-20\,%
\operatorname{nm}%
$ and find that the angular dependent STT falls into the experimental
range\cite{Ripp2010} estimated by Slonczewski's formula for symmetric spin
valves with $\Lambda=1.3$ and $1.7$. Experimentally, Co$_{90}$Fe$_{10}$ is
used as fixed layer. Its spin-flip diffusion length is shorter than that of
Co, but its resistivity is also higher, so there is not much difference when
compared with a Co polarizer. We assume that the interface is not affected. We
plot the results of CoFe in Fig. \ref{tor_comp} with $d_{\mathrm{Co}%
_{90}\mathrm{Fe}_{10}}=2.5\,%
\operatorname{nm}%
$ (Ref.
\onlinecite{Bass1999}%
), and $\rho_{\mathrm{Co}_{90}\mathrm{Fe}_{10}}=154\,%
\operatorname{\Omega }%
\operatorname{nm}%
$ and is very similar to pure Co with $d_{\mathrm{Co}}=5\,%
\operatorname{nm}%
$. The results for Co$|$Cu$|$(Co$_{1}$Ni$_{x}$)$_{y}$Co$_{1}|$Cu from $x=2-3$
and $y=2-5$ are shown in Fig. \ref{coni}. We observe large difference between
epitaxial and disordered samples, but only weak dependences on $x\ $and $y$.
The results for epitaxial (disordered) samples fall into the range of
Slonczewski's $\Lambda=1.05-1.15$ ($1.4-1.5)$.

The experimental results were parameterized by Eq. (\ref{TORQUE}) for a
structurally symmetric spin valve, whereas our results are based on the theory
for asymmetric structures.\cite{Alex2002} We suggest that in future
experiments Slonczewski's formula should be replaced by a more accurate parameterization.

Another interesting material with perpendicular magnetic anisotropy is
Co$|$Ru. Experimentally, both hcp(0001)\cite{Gabaly2007} and
fcc(111)\cite{Eid2002} structures have been reported. Despite the large
lattice mismatch between Co and Ru, hcp Co$|$Ru could be grown epitaxially and
the magnetic anisotropy depends on the thickness of the Co
layer.\cite{Liu1992} However, the metastable structure relaxes to a more
stable one after annealing.\cite{Liu1992} Co$|$Ru$|$Co with a metastable
fcc(111) structure has also been reported.\cite{Bloemen94} Here we present
systematic calculations of the transport properties of Co$|$Ru pillars with
different structure and lattice constants as listed in Table \ref{comp}.

For epitaxial samples, we show results for an fcc(111) texture with lattice
parameters for Ru, Co, and its average. The lattice parameter along the growth
direction is varied to keep the atomic volume constant. Both spin polarization
and specific resistance are close to the experimental values,\cite{Eid2002}
but considering the large lattice distortion ($7.3\sim14\%$) this may be accidental.

For the epitaxial hcp(0001) texture our calculations yield very high spin
polarizations $\tilde{P}=-39\sim-55\%$ for both clean and dirty interfaces
when Co adopts the Ru structure and lattice constants as
reported.\cite{Rahmouni99} and small specific resistances $A\tilde{R}%
=A/\tilde{G}=0.69-0.78\times10^{-15}%
\operatorname{\Omega }%
\operatorname{m}%
^{2}$. Here and below $\tilde{P}$ and $\tilde{G}$ have been Schep corrected
with magnetically active layer thickness $d_{\mathrm{Co}}=5%
\operatorname{nm}%
$\textit{. }Note that the structure is metastable and under annealing Co is
expected to return to its normal lattice parameter.

To simulate sputtering conditions, a $14\times14$ Co is matched to a
$13\times13$ Ru lateral super-cell for both fcc(111) and hcp(0001), leading to
a spin polarization of $\tilde{P}=-15\%$ and specific resistance of
$AR=0.75\times10^{-15}\,%
\operatorname{\Omega }%
\operatorname{m}%
^{2}$ for a clean fcc(111) texture, $\tilde{P}=-28\%$ and $A\tilde
{R}=0.93\times10^{-15}\,%
\operatorname{\Omega }%
\operatorname{m}%
^{2}$ for a clean hcp(0001) texture. A 50\%-50\% interface alloy has little
effect on fcc(111) texture, but leads to a reduced $\tilde{P}=-19\%$ for the
hcp(0001) texture. The measured spin polarization for the Co$|$Ru interface is
$\tilde{P}=-20\%$ with specific resistance $A\tilde{R}=0.5\times10^{-15}\,%
\operatorname{\Omega }%
\operatorname{m}%
^{2}$ (Ref.
\onlinecite{Eid2002)}%
).

In Table \ref{tab_parametrs} we observe that in contrast to the Co$|$Cu
interface, Co$|$Ru has a negative spin polarization for both fcc(111) and
hcp(0001) orientations. Interesting is the relatively large dimensionless
mixing conductance $\tilde{\eta}$. The predicted very large mixing conductance
implies a large skewness of the angular dependent STT, which makes this
material promising for applications in high-frequency generators.

\begin{figure}[ptb]
\includegraphics[width=8.6cm]{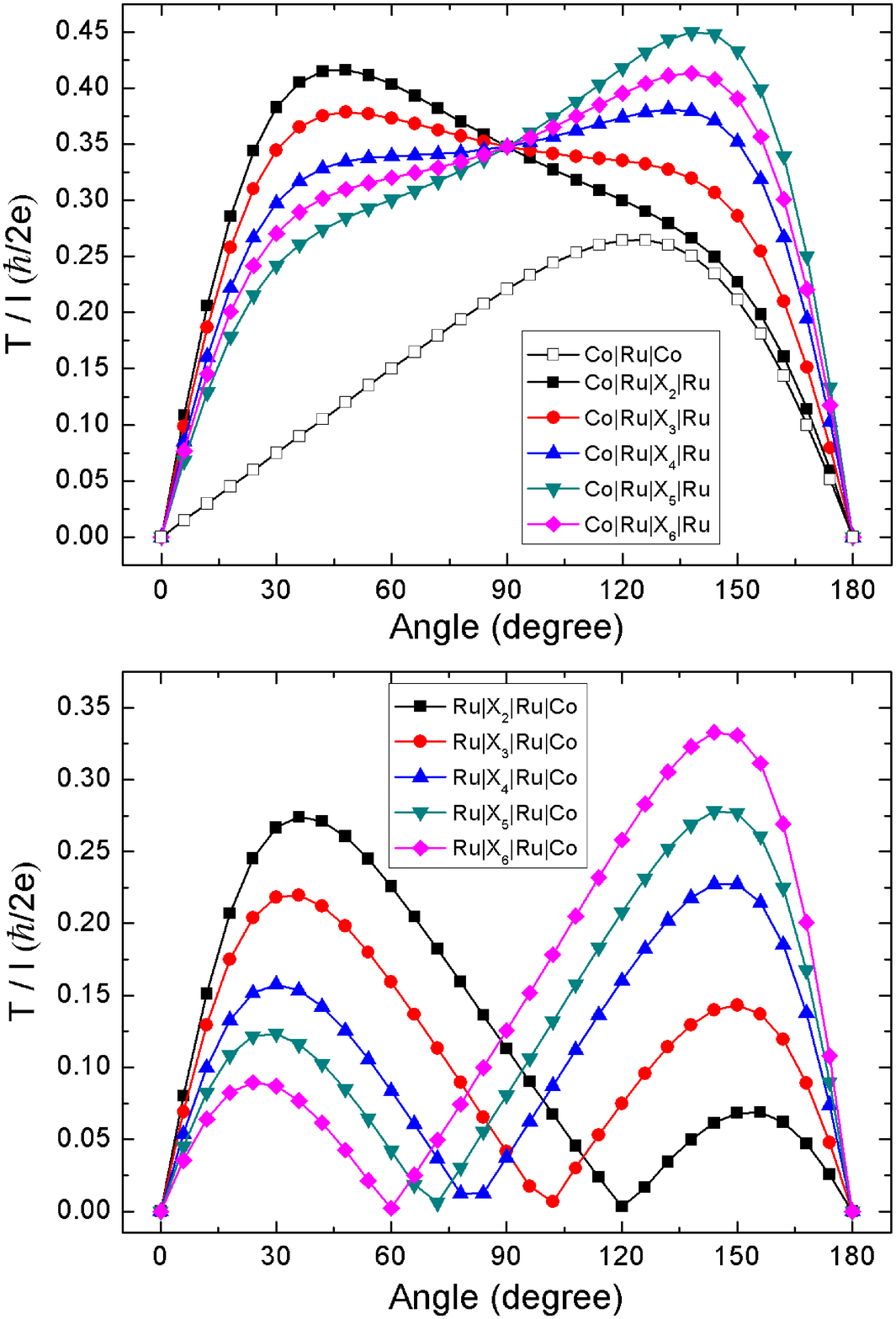} \ \caption{Angular-dependent torkance
$T/I$ on the right-side ferromagnet in disordered Ru based spin valves with
X$_{n}$=(Co$_{1}$Ni$_{2}$)$_{n}$Co$_{1}$. The Co and Ru buffer layers are
assumed much thicker than the spin diffusion lengths $l_{sd}^{\mathrm{Co}}$
and $l_{sd}^{\mathrm{Ru}}$, so that the Schep correction includes the bulk
scattering of the latter length scales. We disregard the bulk scattering in
the $\mathrm{Ru}$ spacer, which should be allowed for the small thickness of 8
monolayer (2.21$\,\operatorname{nm}$) considered.}%
\label{RuCo_torque}%
\end{figure}

Fig. \ref{RuCo_torque} gives the angular dependent $T/I$ in Co$|$Ru$|$%
FM$|$Ru(111) spin valves. Here, disorder is modeled again by two monolayers of
a $\mathrm{50\%-50\%}$ interface alloy (to rather small effect) and Schep and
magnetic bulk corrections have been implemented. When fitted by Slonczewski's
formula, the STT\ on the soft Co$_{1}$Ni$_{2}$ multilayer in the strongly
asymmetric spin valve Co$|$Ru$|\left(  \mathrm{Co}_{1}\mathrm{Ni}_{2}\right)
_{y}$Co$_{1}|$Ru(111) shows a large variation in skewness in terms of the
parameter $\Lambda=0.5-2.0$. The maximum of the angular dependent spin torque
is shifted gradually from low angle to high angle when the thickness of
Co$_{1}$Ni$_{2}$ increases from 2 to 6 periods. When Co serves as the free
layer, the (modulus of the) angular dependent torkance shows two peaks and a
compensation point when the thickness of Co$_{1}$Ni$_{2}$ increases from 3 to
5 periods. This shape can be understood in terms of the spin accumulation in
the normal metal spacer in the parallel configuration,\cite{Manschot} which is
accompanied by a non-monotonic angular magnetoresistance. This behavior has
been observed in Py$|$Cu$|$Co and dubbed `wavy torques'.\cite{Wavy,Fert2008}

\section{Summary}

\label{sec:sum}

We studied the angular dependent STT for materials with
magnetization normal to the interfaces by circuit theory in
combination with first-principles
calculations. An interesting angular dependent STT is found in Co$|$%
Ru$|$(Co$_{1}$Ni$_{2}$)$_{x}$Co$_{1}|$Ru(111) spin valve. Moreover,
a `wavy'
angular-dependent STT acts on the Co layer in Co$|$Ru$|$(Co$_{1}$Ni$_{2}%
$)$_{x}$Co$_{1}|$Ru(111) structures. When the CoNi is the free
layer, we expect very efficient high-frequency generation.

\begin{table*}[ptb]
\begin{center}%
\begin{tabular*}
{17.5cm}[c]{@{\extracolsep{\fill}}ccccccc}\hline\hline system &
$G_{\uparrow}$ & $G_{\downarrow}$ & $\operatorname{Re}G_{\uparrow
\downarrow}$ & $\operatorname{Im}G_{\uparrow\downarrow}$ &
$\tilde{P}$ & $\tilde{\eta}$\\\hline Cu$|$X$_{2}|$Cu & 0.41(0.41) &
0.35(0.19) & 0.55(0.54) & -0.02(-0.03) &
0.25(0.69) & 0.85(1.1)\\
Cu$|$X$_{3}|$Cu & 0.41(0.41) & 0.32(0.18) & 0.56(0.54) &
-0.03(-0.03) &
0.36(0.72) & 0.96(1.1)\\
Cu$|$X$_{4}|$Cu & 0.41(0.41) & 0.31(0.16) & 0.56(0.54) &
-0.03(-0.03) &
0.39(0.75) & 0.98(1.2)\\
Cu$|$X$_{5}|$Cu & 0.41(0.41) & 0.30(0.15) & 0.55(0.54) &
-0.03(-0.03) & 0.42(0.77) & 0.97(1.2)\\\hline Cu$|$Y$_{2}|$Cu &
0.39(0.40) & 0.30(0.21) & 0.40(0.54) & -0.02(-0.03) &
0.34(0.62) & 0.62(1.2)\\
Cu$|$Y$_{3}|$Cu & 0.39(0.39) & 0.26(0.19) & 0.39(0.54) &
-0.02(-0.03) &
0.46(0.64) & 0.66(1.3)\\
Cu$|$Y$_{4}|$Cu & 0.39(0.39) & 0.24(0.17) & 0.40(0.54) &
-0.01(-0.03) & 0.52(0.69) & 0.71(1.3)\\\hline Ru$|$Co & 0.32(0.29) &
0.58(0.53) & 0.92(0.88) & 0.001(0.02) & -0.15(-0.17) &
8.9(8.7)\\
Ru$|$X$_{2}|$Ru & 0.25(0.25) & 0.36(0.31) & 1.03(0.94) & -0.02(0.02)
&
-0.26(-0.15) & 4.8(4.6)\\
Ru$|$X$_{3}|$Ru & 0.25(0.25) & 0.35(0.27) & 1.03(0.94) & -0.02(0.02)
&
-0.24(-0.05) & 4.9(5.1)\\
Ru$|$X$_{4}|$Ru & 0.25(0.25) & 0.33(0.23) & 1.03(0.94) & -0.02(0.02)
&
-0.19(0.06) & 5.2(5.7)\\
Ru$|$X$_{5}|$Ru & 0.25(0.25) & 0.31(0.22) & 1.03(0.94) & -0.02(0.02)
&
-0.15(0.08) & 5.5(5.8)\\
Ru$|$X$_{6}|$Ru & 0.25(0.25) & 0.32(0.20) & 1.03(0.94) & -0.02(0.02)
&
-0.19(0.14) & 5.2(6.2)\\
hex-Ru$|$Co & 0.20(0.23) & 0.53(0.32) & 0.83(0.71) & -0.01(0.01) &
-0.28(-0.19) & 10(7.9)\\\hline Cu$|$Co$^{\ast}$ & 0.42(0.42) &
0.36(0.33) & 0.41(0.55) & 0.01(0.03) & 0.51(0.54) &
1.2(2.0)\\\hline\hline
\multicolumn{7}{l}{$^{\ast}$ Ref \onlinecite{xia02}}%
\end{tabular*}
\end{center}
\caption{Parameters for clean (disordered) interfaces (in units of
$10^{15}\,\operatorname{\Omega }^{-1}\operatorname{m}^{-2}$). $\mathrm{X}%
_{n}\mathrm{=(Co}_{1}\mathrm{Ni}_{2}\mathrm{)}_{n}\mathrm{Co_{1};Y}%
_{n}\mathrm{=(Co}_{1}\mathrm{Ni}_{3}\mathrm{)}_{n}$; fcc Cu and Ru
have
Sharvin conductances of $G_{\mathrm{Cu}}^{sh}=0.55\times10^{15}%
\,\operatorname{\Omega }^{-1}\operatorname{m}^{-2}$ and
$G_{\mathrm{Ru}\left(
fcc\right)  }^{sh}=0.98\times10^{15}\,\operatorname{\Omega }^{-1}%
\operatorname{m}^{-2}$, respectively. $G_{\mathrm{Co}(fcc),\uparrow}%
^{sh}=0.47$ and $G_{\mathrm{Co}(fcc),\downarrow}^{sh}=1.09\times
10^{15}\,\operatorname{\Omega }^{-1}\operatorname{m}^{-2}$ for
majority and minority spin in fcc Co, respectively. For hex Ru's
$G_{\mathrm{Ru}\left(
hex\right)  }^{sh}=$ $0.80\times10^{15}\,\operatorname{\Omega }^{-1}%
\operatorname{m}^{-2}$ and Co with hex Ru structure $G_{\mathrm{Co}%
(hex),\uparrow}^{sh}=0.40$ and $G_{\mathrm{Co}(hex),\downarrow}^{sh}%
=0.80\times10^{15}\,\operatorname{\Omega
}^{-1}\operatorname{m}^{-2}$ for majority and minority spins,
respectively. A magnetically active bulk region correction is
implemented for the normalized spin polarization $\tilde{P}$ and
relative mixing conductance $\tilde{\eta}$. For Cu$|$Co interface,
we use $d_{\mathrm{Co}}=5\,\operatorname{nm}$ (Ref.
\onlinecite{Ripp2010}), and bulk resistivity
$\rho_{\mathrm{Co}}=60\,\operatorname{\Omega }\operatorname{nm}$
with spin asymmetry $\beta=0.46$, which results in $\rho_{\mathrm{Co}%
}^{\uparrow}=81\,\operatorname{\Omega }\operatorname{nm}$ and $\rho
_{\mathrm{Co}}^{\downarrow}=219\,\operatorname{\Omega
}\operatorname{nm}$
(Ref.\onlinecite{Bass1999}). We use a spin diffusion length $l_{sd}%
^{\mathrm{Co}}=60\,\operatorname{nm}$.}%
\label{tab_parametrs}%
\end{table*}

\begin{table}[ptb]
\begin{center}%
\begin{tabular*}
{8.5cm}[c]{@{\extracolsep{\fill}}cccc}\hline\hline
system & lattice & $\tilde{P}$ & $AR(10^{-15}\unit{\Omega}\unit{m}^{2}$) \\
\hline
fcc(111) & Ru & -0.12 (-0.05) & 0.56 (0.83)\\
fcc(111) & Co & -0.14 (-0.20) & 0.67 (0.93)\\
fcc(111) & (Ru+Co)/2 & -0.09 (-0.02) & 0.60 (0.87)\\
fcc(111) & Matching$^{\ast}$ & -0.15 (-0.17) & 0.75 (0.86)\\\hline
hcp(0001) & Ru$^{\dag}$ & -0.55 (-0.39) & 0.78 (0.69)\\
hcp(0001) & Matching$^{\ast}$ & -0.28 (-0.19) & 0.93 (0.98)\\\hline
exper.\cite{Eid2002} &  & -0.2 & 0.5\\\hline\hline
\multicolumn{4}{l}{$^{{\ast}}$$14\times14$ Co matched to $13\times13$
Ru\cite{Grimsditch1996,Gabaly2007}}\\
\multicolumn{4}{l}{$^{\dag}$Cobalt's atomic volume expanded to that of Ru.}%
\end{tabular*}
\end{center}
\caption{Comparison of the calculated spin polarization $\tilde{P}$ (we use
$d_{Co}=5\operatorname{nm}$) and specific interface resistances $A\tilde
{R}=A/\tilde{G}$ of clean (disordered) Co$|$Ru for different lattice
parameters with experiment.}%
\label{comp}%
\end{table}

\section{Acknowledgements}

The authors acknowledge Y. Xu and S. Wang for their preliminary
work. The authors gratefully acknowledge financial support from
National Basic Research Program of China (973 Program) under the
grant No. 2011CB921803, and NSF-China grant No. 60825404, the Dutch
FOM Foundation, and EU-ICT-7 contract no. 257159 MACALO.

\end{document}